# Climate change impacts on wind energy potential in the European domain

R. Davy[1], N. Gnatiuk[2], L. Pettersson[1] and L. Bobylev[1,2]

[1] Nansen Environmental and Remote Sensing Centre/Bjerknes Centre for Climate Research, Bergen, Norway
[2] Nansen International Environmental and Remote Sensing Centre, St. Petersburg, Russia

**Abstract**
We may anticipate that climate change will bring changes to the intensity and variability of near surface winds, either through local effects or by altering the large-scale flow. The impact of climate change on European wind resources has been assessed using a single-model-ensemble of the latest regional climate model from the Rossby Centre, RCA4. These simulations used data from five of the global climate models in the contemporary Climate Model Intercomparison Project (CMIP5) as boundary conditions, and the results are publicly available under the COordinated Regional climate Downscaling EXperiment (CORDEX) project. Overall we find a consistent pattern of a decrease in the wind resources over the European domain under both the RCP 4.5 and RCP 8.5 scenarios, although there are some regions, principally North Africa and the Barents Sea, with projected increases in wind resources. The pattern of change is both robust across the choice of scenario, and persistent: there is a very similar pattern of change found in the latter part of the 21$^{st}$ century as in the earlier. A case study was chosen to assess the potential for offshore wind-farms in the Black Sea region. We developed a realistic methodology for extrapolating near-surface wind speeds up to hub-height using a time-varying roughness length, and determined the extractable wind power at hub-height using a realistic model of contemporary wind-turbine energy production. We demonstrate that, unlike much of the Mediterranean basin, there is no robust pattern of a negative climate change impact on wind resources in the studied regions of the Black Sea. Furthermore, the seasonality of wind resources, with a strong peak in the winter, matches well to the seasonality of energy-demand in the region, making offshore wind-farms in the Black Sea region a viable source of energy for neighbouring countries.

## 1. Introduction
The increasing demand for renewable energy sources in the coming decades (IEA, 2013) requires that we have a clear understanding of accessible wind resources, and the susceptibility of these resources to climate change (Schaeffer *et al*., 2012). The wind energy industry has clear ambitions of expansion, with wind energy capacity scenarios ranging from 251 to 392 GW production by 2030 (EWEA, 2015), between two to three times the present (2014) capacity. For a location to be suitable for the construction of a wind farm it must have sufficiently high and reliable wind speeds, and for these to not decrease significantly under future climate change. Wind energy potential is strongly dependent upon the strength of near-surface winds which are determined by synoptic-scale variability and local processes, such as those related to orography. Therefore if climate change alters the large-scale flow (Lu et al., 2007; Mizuta, 2012) or local conditions such as stratification (Wharton and Lundquist, 2012) or surface roughness (Vautard et al., 2010) it can also alter the available wind resources, and indeed it has been shown that this has already happened (Klink, 1999; Smits et al., 2005; McVicar et al., 2008; Vautard et al., 2010). This is especially important in the planning of new wind-farms given the need for the timing of power generation to match the timing of demand, due to the lack of large-scale energy storage (Goddard et al., 2015).

A common tool to explore the potential for the impacts of climate change on wind energy resources is the regional climate model (RCM). These have the advantage over global climate models that, due to the reduced domain, one can readily run these models at a higher spatial resolution and with more detailed surface processes, to better capture changes to the near-surface winds (Samuelsson et al., 2015). The RCM inter-comparison project, EURO-CORDEX (Jacob et al., 2014), has provided an ensemble of historical and future simulations of climate which has been a rich resource for studying the effects of regional climate change on winds (Balog et al., 2016; Tobin et al., 2016). However, there are relatively few studies which have used RCM projections to assess the near-surface wind resources using realistic models of wind power production at the typical heights of wind-turbines. This is partly due to the relatively high computational cost of applying realistic power-curve models of extracted wind power as a function of wind speed, and due to the complexities of extrapolating the typical model output of wind speeds at 10 m above the surface, to wind speeds at hub-height (Schaeffer et al., 2012). However, both of these factors are important in assessing wind resources due to the dependency of the extrapolation of 10 m wind speeds to hub-heights on local conditions, such as roughness length (Stull, 2012; Mirhosseini et al., 2011), and because a realistic model of extractable wind potential is a primary metric in determining the feasibility of a wind-farm (IEA, 2013). While it is common to use simple power-law approximations to extrapolate wind speeds to turbine hub-heights (Elliot, 1979; Tobin et al., 2016), in this work we developed a roughness-length dependency in our calculation of wind speeds based on established boundary-layer profiles for winds (Stull, 2012).

In this study we have performed an ensemble analysis of the Rossby Centre's latest regional climate model, RCA4 (Strandberg et al, 2014) set up for Europe at an approximately 12 km (0.11° on a rotated grid) spatial resolution. Others have assessed the previous iteration of this regional climate model (RCA3) in terms of the reproducibility of wind resources in localized regions (Nolan et al., 2012; Pryor et al., 2012), and have demonstrated that the model performed well in reproducing the current wind speed climatology. There are three main sources of uncertainty associated with regional climate model assessments: the choice of global climate model used for the boundary conditions; the choice of regional climate model; and the internal variability of the climate models. An ensemble of RCA4 simulations of European climate is a useful way to assess the robustness of signals of climate change in the wind resources as resolved in the regional model projections. It allows us to quantify the sensitivity of simulated wind resources to the choice of global climate model used to prescribe the boundary conditions. However, it has been shown that the uncertainty in wind speeds that comes from the choice of regional climate model may be comparable to, or greater than that which comes from either the choice of global model or the internal variability (Outten and Esau, 2013; Tobin et al., 2015). Therefore we need to exercise some caution when interpreting results from such a single-model-ensemble of climate projections as used here.

Here we use an ensemble of climate projections to assess the susceptibility of wind resources in the European domain to climate change, with a more in-depth assessment of the potential for offshore wind-farms in the Black Sea region. In Section 2 we discuss the datasets and methodologies used in this paper; in Section 3 we review the single model ensemble for consistency with a contemporary reanalysis product, ERA-Interim; in Section 4 we assess the wind resources in the European domain and the projected changes in the 21$^{st}$ century; in Section 5 we take a case study on the potential for wind farms in the Black Sea region and their susceptibility to climate change; and finally in Section 6 we present our conclusions.

## 2. Data and Methods

The regional climate model data for the RCA4 CORDEX simulations were taken from publicly available Swedish Meteorological and Hydrological Institute (SMHI) archives (Strandberg et al, 2014). The five different global climate models used to drive the regional climate model are detailed in Table 1. From this archive of regional simulations we obtained the 10m wind speeds, $U_{10}$ and $V_{10}$, at a 3-hourly resolution; and the surface pressure ($p_s$), surface air temperature ($T_s$), components of the surface wind stress: $\tau_U$ and $\tau_V$, and the components of the wind speed at 850hPa: $U_{850}$ and $V_{850}$, at the daily resolution for the EUR-11 domain from the historical, RCP 4.5 and RCP 8.5 scenarios used in CMIP5. From these scenarios we selected three periods of interest: a period henceforth referred to as 'historical', 1979-2004; 'Future1' which covers the early 21$^{st}$ century, 2021-2050; and 'Future2' for the later 21$^{st}$ century, 2061-2090. For the historical period we also acquired $U_{850}$, $V_{850}$, $U_{10}$ and $V_{10}$ wind speed data of the ERA-Interim reanalysis at a 0.125$^{o}$ x 0.125$^{o}$ resolution from the European Centre for Medium range Weather Forecasting archives.

| Origin of global climate model | Acronym |
|---|---|
| Institut Pierre Simon Laplace, Paris, France | IPSL-CM5A |
| Centre of basic and applied research (CERFACS)/Météo-France/ National centre for scientific research (CNRS), Toulouse, France | CNRM-CM5 |
| EC-EARTH consortium of national weather services and universities | EC-EARTH |
| UK Met Office Hadley Centre, Exeter, UK | HadGEM2-ES |
| Max Planck Institute for Meteorology, Hamburg, Germany | MPI-ESM-LR |

**Table 1.** List of the producers of the different global climate models (from CMIP5) that were used for boundary conditions for the CORDEX regional simulations used here, and the acronyms we use in this manuscript.

The 3-hourly resolution wind speed was averaged to daily resolution for the purposes of calculating the wind speeds at a height of 120m above the surface – the typical hub-height of a wind turbine. In order to extrapolate the wind speeds from the available data, given at a height of 10 m above the surface, we employ the logarithmic profile for the wind speed (Stull, 2012). This is given by the relationship:

$$W_z = \frac{u_*}{k} \ln \frac{z}{z_0} \quad \text{(Eq. 1)}$$

where $W_z$ is the wind speed at height $z$; $k$ is the von Karman constant, taken to be 0.4; $z_0$ is the roughness length; and $u_*$ is the friction velocity. The friction velocity is calculated from the surface wind stress ($\tau$) and the air density ($\rho$):

$$u_* = \sqrt{\tau/\rho} \quad \text{(Eq. 2)}$$

where $\tau = \sqrt{\tau_U^2 + \tau_V^2}$, and we take $\rho = p_s/RT_s$, with $R$=287.06 J kg$^{-1}$ K$^{-1}$.

First we calculated the friction velocity from the available data using Eq. 2, and then used the 10 m wind speeds to calculate the roughness length using Eq. 1. We enforced a lower limit on the roughness length, based on values found over calm water, such that $z_0 \geq 0.0002$ m (Stull, 2012). The derived friction velocity and roughness length were then substituted back into Eq.1 to obtain the wind speed at a height of 120 m above the surface – the turbine height. Note that equation 1 includes an implicit assumption that the atmosphere is neutrally-stratified, and in principle a term accounting for the stability of the atmosphere should be included. However, this was not possible here given the limited availability of data. So when we use this formulation to estimate wind speeds at 120 m, we will tend to over-estimate wind speeds in convective conditions and under-estimate wind speeds in stably-stratified conditions. Unfortunately, without having information on the climatology of the stratification, we cannot assess how this will affect our results.

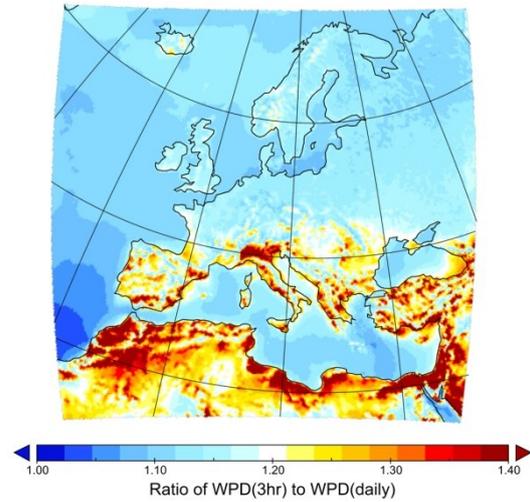

Figure 1. The ratio of WPD calculated using 3-hourly resolution and daily resolution wind speeds, averaged over the Historical period of the downscaling of CNRM-CM5 data.

The Wind Power Density (WPD), the power per unit area, is defined from the air density and the wind speed:

$$WPD = \frac{P}{A} = \frac{1}{2}\overline{\rho W^3} \quad \text{(Eq. 3)}$$

where $W$ is the wind speed at a height of 120 m and the over-bar indicates an averaging over time. The calculated WPD is very sensitive to the small-timescale variations in wind speed because of this cubed dependence on wind speed. Ideally we would use wind speed data at a time resolution consistent with the response-time of wind turbines i.e. 10-minute averages. However, here we are constrained to using daily-mean wind speed data, and so we anticipate that we will under-estimate the wind power density. We have tested this limitation using the available 3 hourly resolution 10m wind speed data. We calculated the WPD at a height of 10 m using the daily data, and an alternative WPD by multiplying the cubed wind speed in Eqn. 3 by the ratio of the cubed wind speeds at 3-hourly and daily resolution ($W_{3hr}^3/W_{daily}^3$) at each day (Figure 1). The biggest differences in the two WPD estimates are seen over land in the Mediterranean region, with generally less than 20% differences over water, although there are some strong geographical variations.

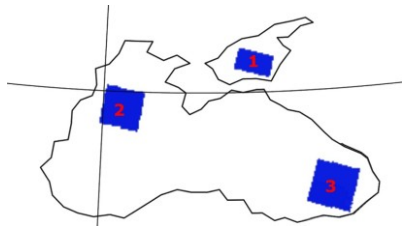
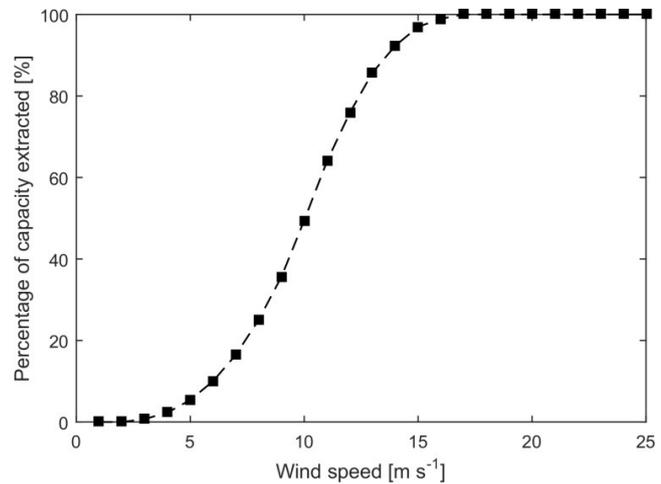

**Figure 2. Left:** The three regions for which we assess the extractable Wind Power: (1) Azov Sea, (2) NW Black Sea, and (3) SE Black Sea. **Right:** The power-curve of the Enercon E-126 wind turbine, expressed in percentage of the maximum capacity of the turbine (Enercon, 2015).

To calculate the Extractable Wind Power (EWP), given the current level of technology, we used power-curve data from a typical 120m hub-height wind turbine. The data is from the high capacity, 7.6 MW Enercon E-126 turbine (Enercon, 2015) and its normalized power curve is given in Figure 2. We used a spline interpolation to calculate the Wind Power produced at each time-step, which was then averaged over the period of interest to obtain the typical EWP for a given location. We chose three locations in the Black Sea region for analyzing the EWP, highlighted in Figure 2, and henceforth referred to as Azov Sea, NW Black Sea, and SE Black Sea.

When correlations are reported, these are spatial correlations tested for significance using a two-tailed Student-t distribution.

## 3. Current representation of winds in RCA4

In order to assess the ability of the RCA4 regional climate model to characterize the current wind resources in Europe we compare the climatology of the wind speeds in the historical ensemble to those found in an independent reanalysis product, ERA-Interim. We analyze both the consistency in the large-scale flow, as characterized by the wind speed and direction at 850hPa, and the near surface winds at a height of 10m for our area of interest, the Black Sea.

There is a broad consistency between the ensemble-mean wind speeds and direction over the European domain (R=0.88, p<0.01), although there is a general bias towards lower wind speeds in the model-ensemble over Northern Europe, especially in the storm track between Scotland and Iceland (Figure 3a). While the wind speeds over the Mediterranean and Black Sea regions are quite consistent between the model-ensemble and ERA-Interim, the prevailing wind direction is rather different. This can be seen clearly in the Black Sea region (Figure 3b) where the model-ensemble and reanalysis have close agreement in the Western part of the domain, but in the model-ensemble the winds have a westerly bias in the West of the Black

Sea. Overall we see greater wind speeds in the model ensemble than in the reanalysis in this region, which indicates that the projections can overestimate the wind power density here.

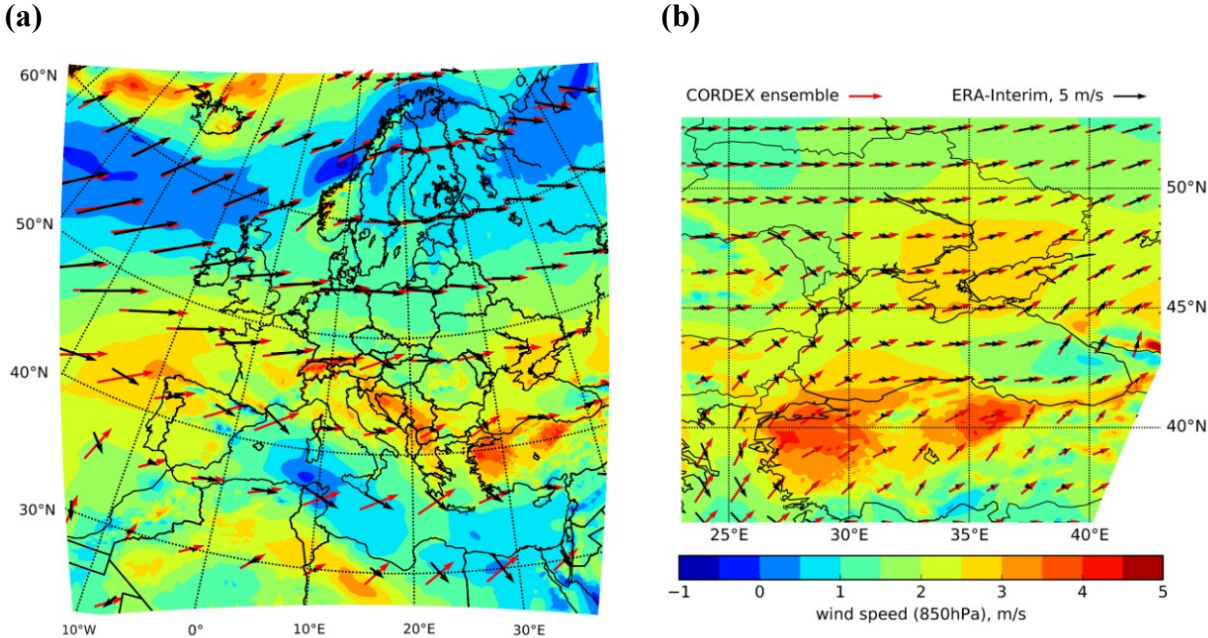

**Figure 3.** The difference between the RCA4 model ensemble-mean and ERA-Interim in the climatological-mean wind speed and direction at a height of 850 hPa, for (a) the European domain and (b) the Black Sea region

Naturally the higher resolution of RCA4, compared to the ECMWF model used to create ERA-Interim, leads to a better representation of the difference in near-surface wind speeds between the land and water, as can be seen in Figure 4a, 4b. There is also a general bias in the models to produce higher wind speeds over the Black Sea region than is found in the reanalysis, which is consistent with the difference in upper-level winds and the expectation that the differences in surface wind speeds are driven by the synoptic conditions. Furthermore, the overall pattern of stronger winds over the NW Black Sea and Azov Sea regions, decreasing towards the SE Black Sea is consistent in both the model ensemble and the reanalysis.

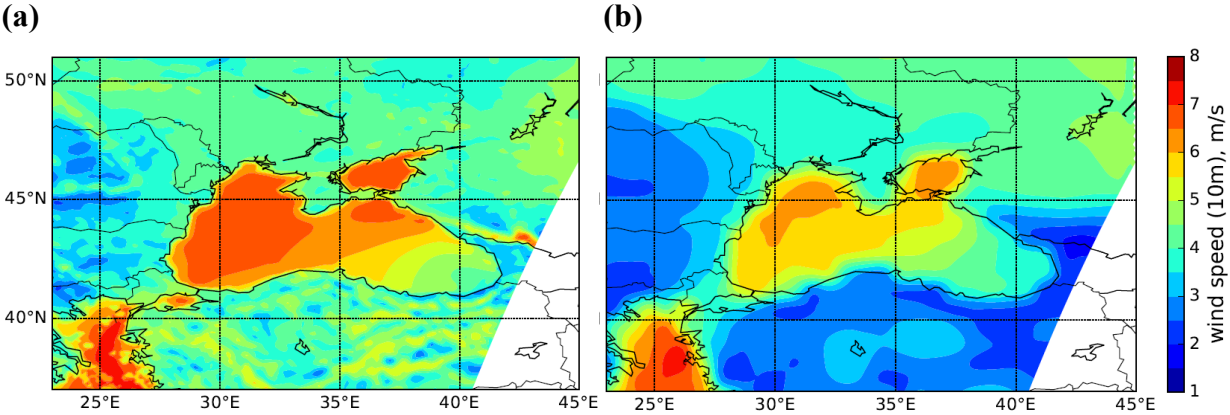

**Figure 4.** The mean wind speed at a height of 10 m above the ground, from (a) the RCA4 model ensemble-mean and from (b) ERA-Interim reanalysis.

As well as the biases in the mean surface wind speeds, it is also necessary to compare the variability of the surface winds in the RCA4 simulations and the ERA-Interim reanalysis. Figure 5 shows Taylor plots (Taylor 2001) comparing the different, downscaled global climate models to ERA-Interim for the monthly-mean wind speeds and monthly-anomaly wind speeds, averaged over the three selected Black sea locations. In the plots of the monthly-mean wind speeds we can see that the RCA4 model results closely match the ERA-Interim reanalysis variability over the Azov Sea and the NW Black Sea, but are systematically biased towards over-estimating variability in the SE Black Sea. The corresponding Taylor diagrams of monthly anomalies in wind speed show that much of this bias in the wind speed variability comes from a higher inter-annual variability in the models, as compared to the reanalysis.

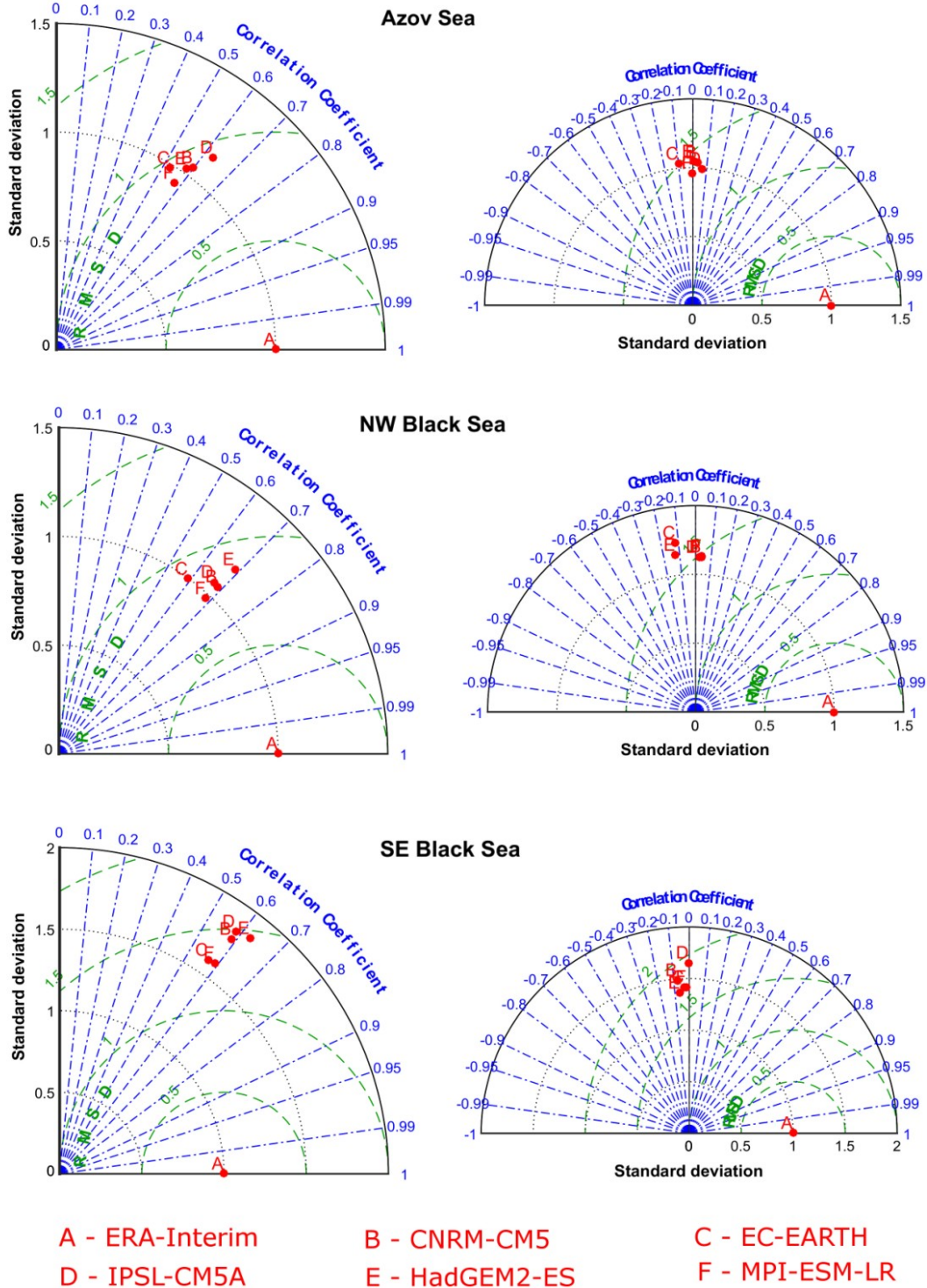

**Figure 5**. Taylor diagrams for the monthly-mean wind speeds (upper) and monthly wind anomalies (lower) comparing each of the global-model downscalings with ERA-Interim in our three regions of interest. The variability (standard deviations) in each of the models has been normalized to the value for ERA-Interim.

## 4. Wind resources and climate change

A reasonable depiction of the mean and variability of the wind speeds in a regional climate model is critical for their use in the assessment of wind resources, because the wind power density (WPD) is proportional to the cube of the wind speed (Eq. 3). Therefore, even small biases in the mean or the variability of the wind speeds can lead to erroneous conclusions about the available wind resources. The geographical distribution of wind power density is closely tied to that of the mean wind speed. For Europe the highest WPD is in the north Atlantic, relatively low values over non-mountainous land, and higher values over open water such as the Mediterranean and the Black Sea (Figure 6). In our case-study region of the Black Sea we find that the present (1979-2004) wind resources are good (WPD>500 W m$^{-2}$) in the Azov Sea region, moderate resources (500 W m$^{-2}$>WPD>300 W m$^{-2}$) in the North and West of the Black Sea, and poor power density (WPD<300 W m$^{-2}$) in the South-Eastern Black Sea (Figure 6b).

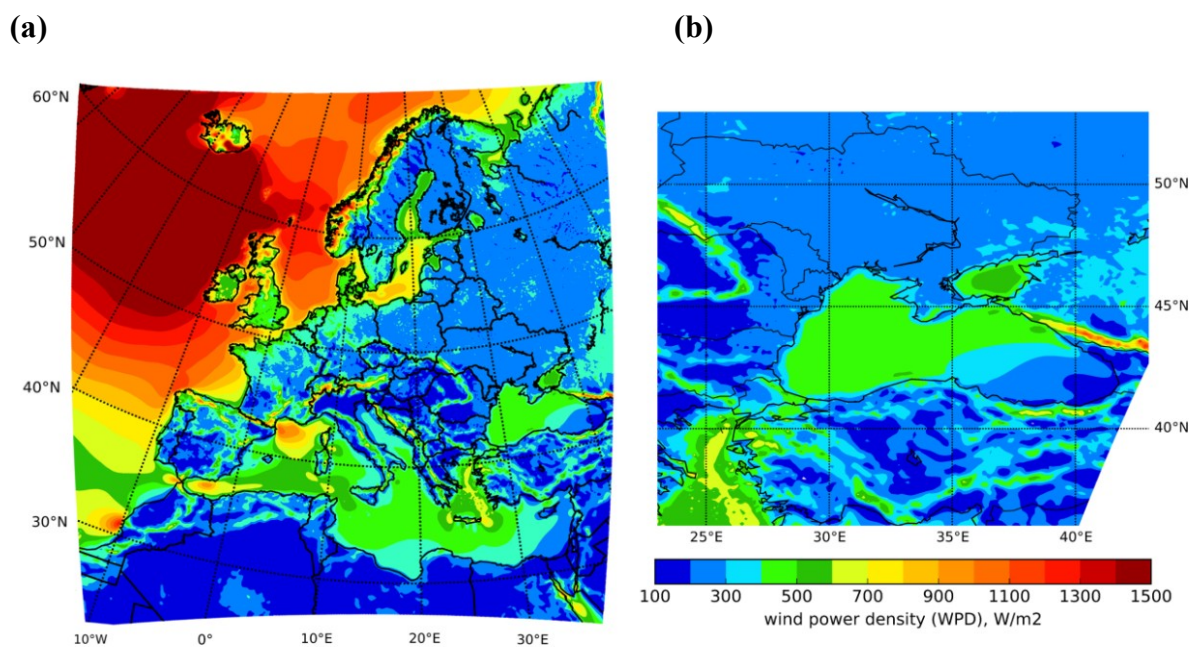

**Figure 6.** The climatological mean Wind Power Density in (a) the European domain and (b) the Black Sea region from the ensemble-mean CORDEX simulations over the historical period (1979-2004).

The projected changes in Wind Power Density between the historical and Future1 period, and between the historical and Future2 period, are shown in Figure 7. The changes are shown only where at least four of the five simulations agree as to the direction of change in the WPD (henceforth referred to as a robust signal). The geographical pattern of changes is consistent between the RCP 4.5 and RCP 8.5 scenarios (R=0.94, p<0.01 in Future1, R=0.96, p<0.01 in Future2), although there is a greater magnitude of change in the RCP 8.5 scenario as may be expected given the greater increase in the global temperatures in this scenario. The climate change signal is also persistent with a strong similarity between the pattern of change from the historical period to the Future1 as from the historical to Future2, i.e. the geographical pattern of change is persistent throughout the 21$^{st}$ century (R=0.90, p<0.01 in RCP 8.5, R=0.94, p<0.01 in RCP 4.5). There is a general decrease in WPD across most of the European domain with only a few locations, including the Baltic Sea, Barents Sea and the interior of North Africa, showing a robust signal of increased WPD. This is largely consistent with the findings from the ENSEMBLES dataset (Tobin *et al.*, 2015).

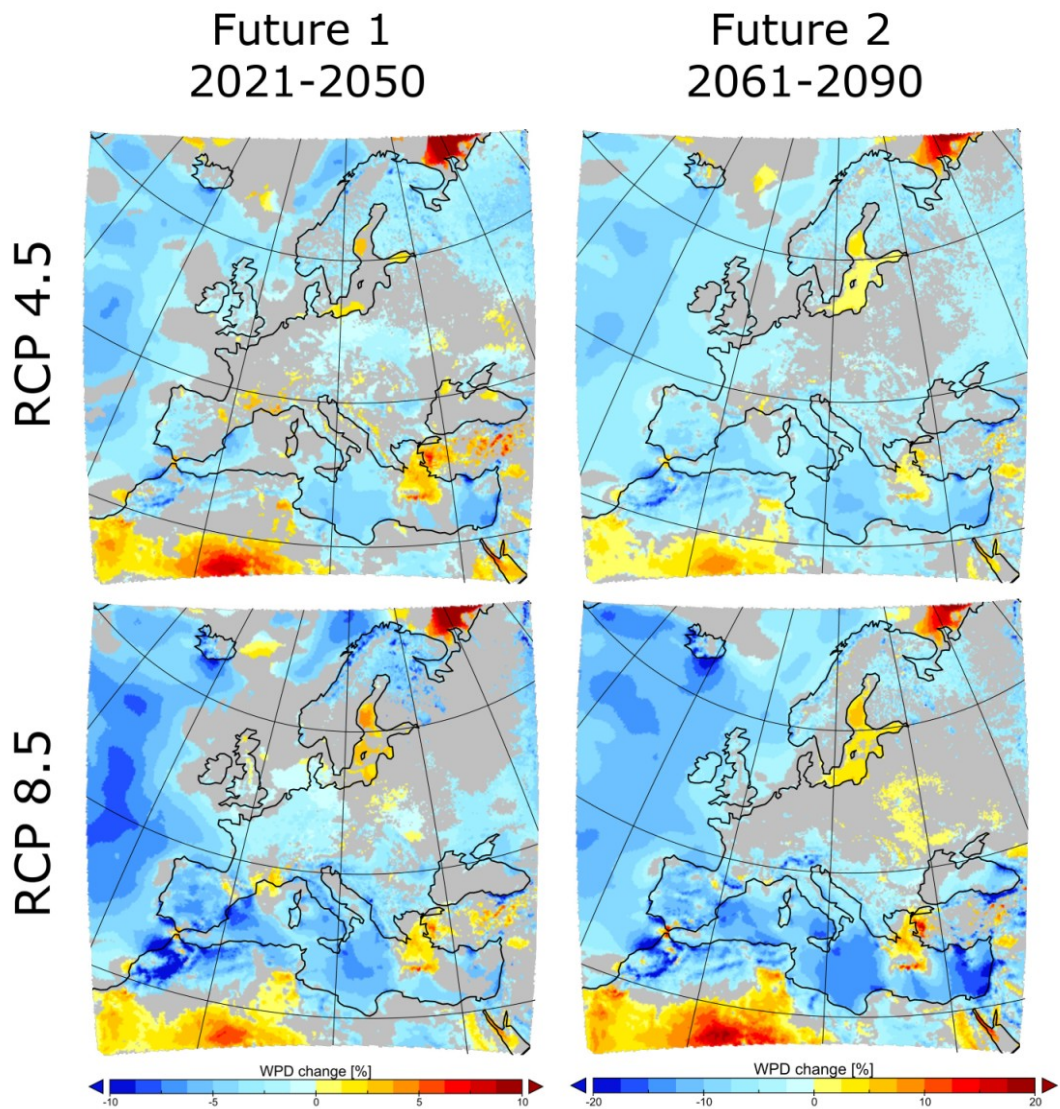

**Figure 7.** The percentage change in Wind Power Density from the historical period to the Future1 period (left column) and to the Future2 period (right column), under the RCP 4.5 (upper) and RCP 8.5 (lower) scenarios. This is the ensemble-mean change, shown for all regions where at least four of the five simulations agreed as to the direction of change (other areas are marked in grey).

**5. Wind energy in the Black Sea**

While assessing changes to the Wind Power Density is useful in describing the potential wind resources, when it comes to planning a wind-farm it is more useful to know how the Extractable Wind Power will be affected by climate change. In practice the full wind power density cannot be extracted by wind turbines, due to limitations from fluid dynamics (the Betz limit (Van Kuik, 2007)), and due to the technology: wind turbines need a minimum wind speed in order to begin operations, and they have a cut-out speed at which they cease functioning in order to prevent damage. The fraction of power that a wind turbine can extract from the wind also varies within the range of wind speeds in which they operate. This can be described by the power curve of a wind turbine: the fraction of peak output that the wind turbine produces at a given wind speed. The example power curve that we use in our calculations is for a typical 120m hub-height turbine, and it is shown in the methods section, Figure 2.

Generally we find a similar dependency on period and scenario for extractable wind power as we do for the wind power density (Figure 8). This indicates that much of the projected change to the wind climatology occurs in the range of operation of the wind turbine. There are few robust patterns of the effect of climate change on the EWP in the Black Sea region, defined as when 4 of 5 model experiments (Table 1) agree as to the sign of change under either of the two warming scenarios used. For example, there is a robust pattern of decreasing EWP in the Azov Sea in the RCP 8.5 scenario in both Future1 and Future2 (even though an individual model may predict a decrease in one period and an increase in another period), while there is no such agreement in the change of WPD in the Azov sea. There are no robust patterns of change in either WPD or EWP in the Azov Sea under the RCP 4.5 scenario. In the NW Black Sea there are no robust patterns of changes to EWP or WPD under either RCP scenario. In the SE Black Sea there is a robust pattern of reduced EWP in the later part of the 21$^{st}$ century under RCP 8.5. However, overall the projected changes to the EWP in all cases are small (<10% of the historical EWP), and the differences between periods are comparable to the differences between the different GCM-RCM combinations.

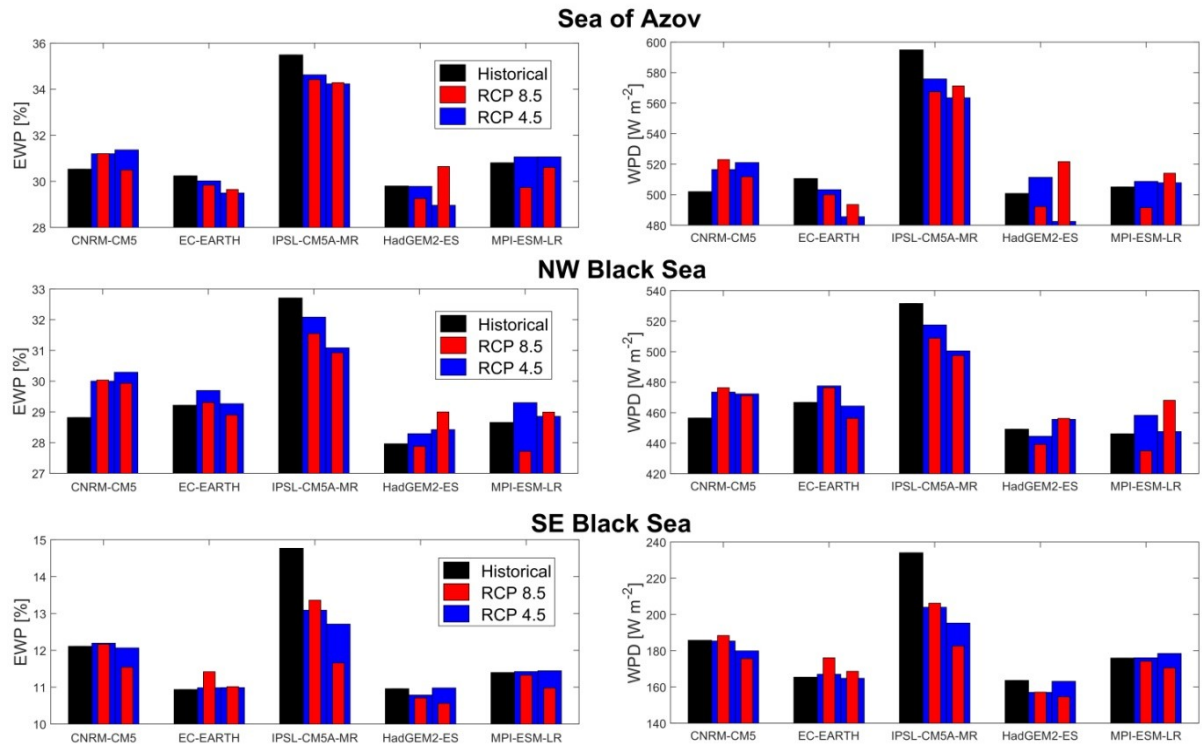

**Figure 8.** The Extractable Wind Power (EWP) in percentage of capacity (left panels) and Wind Power Density (WPD) in W m$^{-2}$ (right panels) for each of the three regions and each of the five global models used for boundary conditions. Results are shown for the historical (left

column), Future1 (middle column) and Future2 (right column) for the historical (black), RCP 4.5 (blue) and RCP 8.5 (red) scenarios.

In considering the potential for wind energy to contribute to national energy supplies, we need to also consider how the seasonal variability in the resource correlates with the seasonality in the demand for energy. There is a strong seasonal cycle to the wind resources in the Black Sea regions (Figure 9), but the analyzed data and simulations did not show a detectable change to these seasonal cycles under climate change (not shown). A mean extractable wind potential fraction greater than 0.3 is generally required to consider a wind-farm to be viable (IEA, 2013). In the two regions with relatively good wind resources, the Azov Sea and NW Black Sea, this is only exceeded in the winter months from September/October to April. However, this seasonal cycle in wind energy production is closely tied to the seasonal cycle of demand for energy resources in this region, due to the primary use of energy for heating (IEA, 2012). Therefore wind energy may very well be a viable resource for countries surrounding the north and west of the Black Sea, such as Ukraine and Russia.

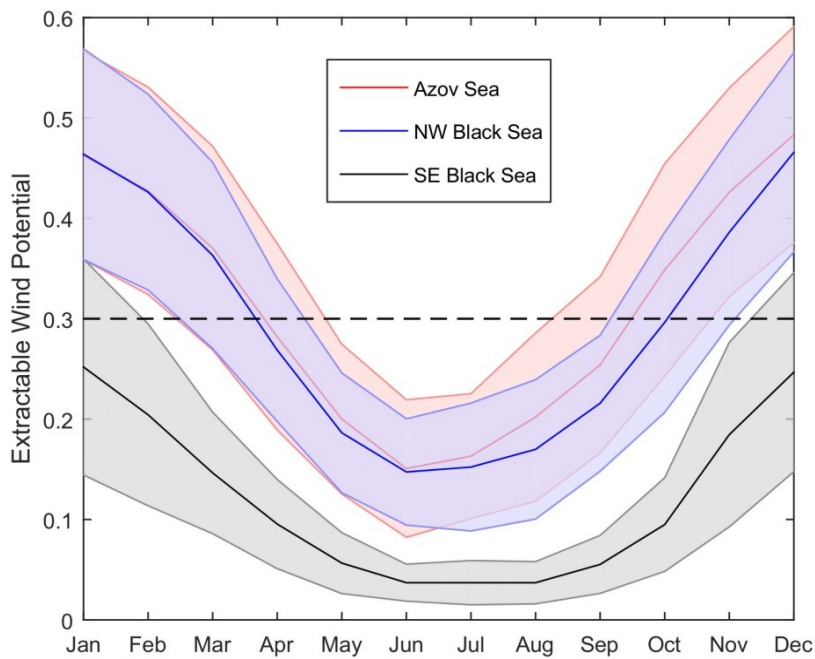

**Figure 9.** The seasonal cycle in the Extractable Wind Potential (EWP) at each of the three sites in the Black Sea from the historical period (1979-2004). The mean is given by the thick lines and the shaded area indicates one standard deviation of the inter-annual variability. The typical threshold value for EWP for consideration of an operational wind farm is 0.3, and is indicated by the horizontal dashed line.

## 6. Discussion

The method we developed here to extrapolate wind speeds up to the hub-height of typical wind turbines (~120m) marks a distinct improvement over other methodologies which have used a fixed roughness length (Goddard *et al.*, 2015; Sliz-Szkliniarz and Vogt, 2011), or simple empirical relationships to extrapolate 10m wind speeds (Elliot 1979; Tobin *et al.*,

2015). Some authors even use the 10m wind speeds themselves to determine wind resources despite the highly non-linear relationship between 10m and hub-height wind speeds (Reyers *et* al., 2014; Tobin *et al.*, 2016). However, the limitation of our methodology is that we had to assume a neutrally-stratified atmosphere in our description of the wind speed profile because we had no information about the stability of the atmosphere. We would suggest that in future, regional climate models should include more information about the near-surface wind speeds at say ~100 m, which is closer to the typical hub-height of wind turbines. This would reduce the large uncertainty in wind resources that comes from the extrapolation of 10m wind speeds up to hub-height. Furthermore, due to the strong dependency of wind power density on the variability of the wind (as discussed in the methods section here), we would suggest that a relatively high temporal resolution (e.g. 3-hourly) is used to calculate wind resources.

We have presented an ensemble-simulation of the wind resources in the European domain from the latest regional climate model from SMHI. Here we have created a realistic methodology to extrapolate surface winds to the hub-height of typical wind turbines using a time-variant roughness length derived from model data. By combining this with power-curve data from a contemporary wind turbine we have assessed the potential for future wind-farms in the Black Sea region. The wind resources were assessed for their reliability, persistence under climate change, and matching to the seasonal variation of energy demand in the region. Since the lifetime of a typical wind farm extends over several decades, climate change can be a significant factor in determining if a wind-farm is viable, especially given that the next decades are likely to be a period of intense climate change (EWEA, 2015). Due to the lack of (or very limited) energy storage available for national electricity networks, it is very important that the timing of power generation during the year matches the seasonal demand for energy. We have shown that the seasonal availability of wind resources in the Black Sea region very closely matches to the seasonal demand for energy, which is principally related to the need for heating in winter, so new wind farms in the region are a potentially viable source of energy for neighboring countries.

In our ensemble analysis of the RCA4 model we assessed the change in wind energy resources from a 'historical' period to the middle and late 21$^{st}$ century. This allowed us to remove any systematic biases in the RCA4 model and to quantify the uncertainty in future wind resources that comes from the choice of global climate model used to simulate the future climate. This was repeated for two scenarios for weak (RCP 4.5) and strong (RCP 8.5) warming. There is a robust signal of decreasing wind resources over much of the European domain in the projections for the 21$^{st}$ century, both towards 2050 and 2100. While this result is consistent with expectations from the previous regional climate model inter-comparison project the pattern of change is distinctly different in this ensemble (Tobin *et al.*, 2015). However, there is no discernible negative influence of climate change on wind resources in our study region of the Black Sea under either the RCP4.5 or RCP8.5 warming scenarios. Overall, the uncertainty due to the choice of global climate model used to prescribe the boundary conditions is a greater source of uncertainty in the wind power density and extractable wind potential in the Black Sea than is climate change, as resolved by the various climate models (table 1).

Overall the northern and western parts of the Black Sea region have been shown to be very suitable for the development of new wind farms. The seasonality of the wind resources (with a peak in wind energy potential in winter) fits well to the seasonality of energy demand in the region, and the wind resources in the Black Sea have been shown to be unaffected by the projected climate change of the 21$^{st}$ century, unlike the wind resources in much of the European domain.

**Acknowledgements**
The research leading to these results has received funding from the European Community's Seventh Framework Programme (FP7) under Grant Agreement No. 287844 for the project "Towards COast to COast NETworks of marine protected areas (from the shore to the high and deep sea), coupled with sea-based wind energy potential" (COCONET).. The RCA4 CORDEX simulation data are taken from the SMHI website https://esg-devel.nsc.liu.se/search/nsc_cog_node_devel/. We would like to thank the ECMWF for making available the ERA-Interim reanalysis product at http://apps.ecmwf.int/datasets/.